\documentclass[12pt]{article}
\usepackage{amsmath}
\usepackage{amscd}
\usepackage{amsfonts}
\usepackage{amssymb}
\catcode`\@11\relax
\newif\ify@autoscale \y@autoscaletrue \def\Yautoscale#1{\ifnum #1=0
  \y@autoscalefalse\else\y@autoscaletrue\fi}
\newdimen\y@b@xdim
\newdimen\y@boxdim \y@boxdim=13pt
\def\Yboxdim#1{\y@autoscalefalse\y@boxdim=#1}
\newdimen\y@linethick    \y@linethick=.3pt
\def\Ylinethick#1{\y@linethick=#1}
\newskip\y@interspace \y@interspace=0ex plus 0.3ex
\def\Yinterspace#1{\y@interspace=#1}
\newif\ify@vcenter   \y@vcenterfalse
\def\Yvcentermath#1{\ifnum #1=0 \y@vcenterfalse\else\y@vcentertrue\fi}
\newif\ify@stdtext   \y@stdtextfalse
\def\Ystdtext#1{\ifnum #1=0 \y@stdtextfalse\else\y@stdtexttrue\fi}
\newif\ify@enable@skew   \y@enable@skewfalse
\expandafter\ifx\csname enableskew\endcsname\relax
 \y@enable@skewfalse \else \y@enable@skewtrue\fi
\def\y@vr{\vrule height0.8\y@b@xdim width\y@linethick depth 0.2\y@b@xdim}
\def\y@emptybox{\y@vr\hbox to \y@b@xdim{\hfil}}
\ify@enable@skew
 \def\y@abcbox#1{\if :#1\else
   \y@vr\hbox to \y@b@xdim{\hfil#1\hfil}\fi}
 \def\y@mathabcbox#1{\if :#1\else
   \y@vr\hbox to \y@b@xdim{\hfil$#1$\hfil}\fi}
\else
 \def\y@abcbox#1{\y@vr\hbox to \y@b@xdim{\hfil#1\hfil}}
 \def\y@mathabcbox#1{\y@vr\hbox to \y@b@xdim{\hfil$#1$\hfil}}
\fi
\def\y@setdim{%
  \ify@autoscale%
   \ifvoid1\else\typeout{Package youngtab: box1 not free! Expect an
     error!}\fi%
   \setbox1=\hbox{A}\y@b@xdim=1.6\ht1 \setbox1=\hbox{}\box1%
  \else\y@b@xdim=\y@boxdim \advance\y@b@xdim by -2\y@linethick
  \fi}
\newcount\y@counter
\newif\ify@islastarg
\def\y@lastargtest#1,#2 {\if\space #2 \y@islastargtrue
  \else\y@islastargfalse\fi}
\def\y@emptyboxes#1{\y@counter=#1\loop\ifnum\y@counter>0
  \advance\y@counter by -1 \y@emptybox\repeat}
\def\y@nelineemptyboxes#1{%
  \vbox{%
    \hrule height\y@linethick%
    \hbox{\y@emptyboxes{#1}\y@vr}
    \hrule height\y@linethick}\vskip-\y@linethick}
\def\yng(#1){%
  \y@setdim%
  \hskip\y@interspace%
  \ifmmode\ify@vcenter\vcenter\fi\fi{%
  \y@lastargtest#1,
  \vbox{\offinterlineskip
    \ify@islastarg
     \y@nelineemptyboxes{#1}
    \else
     \y@ungempty(#1)
    \fi}}\hskip\y@interspace}
\def\y@ungempty(#1,#2){%
  \y@nelineemptyboxes{#1}
  \y@lastargtest#2,
  \ify@islastarg
   \y@nelineemptyboxes{#2}
  \else
   \y@ungempty(#2)
  \fi}
\def\y@nelettertest#1#2. {\if\space #2 \y@islastargtrue
  \else\y@islastargfalse\fi}
\def\y@abcboxes#1#2.{%
  \ify@stdtext\y@abcbox#1\else\y@mathabcbox#1\fi%
  \y@nelettertest #2.
  \ify@islastarg\unskip%
   \ify@stdtext\y@abcbox{#2}\else\y@mathabcbox{#2}\fi%
  \else\y@abcboxes#2.\fi}
 \newdimen\y@full@b@xdim
 \newcount\y@m@veright@cnt
\ify@enable@skew
 \def\y@get@m@veright@cnt#1#2.{%
   \if :#1 \advance\y@m@veright@cnt by 1\y@get@m@veright@cnt#2.\fi}
 \let\y@setdim@=\y@setdim
 \def\y@setdim{%
   \y@setdim@ \y@full@b@xdim=\y@b@xdim
   \advance\y@full@b@xdim by 1\y@linethick}
 \def\y@m@veright@ifskew#1{
   \y@m@veright@cnt=0 \y@get@m@veright@cnt#1.
   \moveright \y@m@veright@cnt\y@full@b@xdim}
\else
 \def\y@m@veright@ifskew#1{}
\fi
\def\y@nelineabcboxes#1{%
\if\space #1 \vbox{\hbox{\y@vr}}\else
  \y@nelettertest #1.
  \ify@islastarg
   \y@m@veright@ifskew{#1}
    \vbox{
      \hrule height\y@linethick%
      \hbox{\ify@stdtext\y@abcbox#1\else\y@mathabcbox#1\fi\y@vr}
      \hrule height\y@linethick}\vskip-\y@linethick
  \else
   \y@m@veright@ifskew{#1}
    \vbox{
      \hrule height\y@linethick%
      \hbox{\y@abcboxes #1.\y@vr}%
      \hrule height\y@linethick}\vskip-\y@linethick
  \fi\fi}
\def\young(#1){%
  \y@setdim%
  \hskip\y@interspace%
  \y@lastargtest#1,
  \ifmmode\ify@vcenter\vcenter\fi\fi{%
  \vbox{\offinterlineskip
    \ify@islastarg\y@nelineabcboxes{#1}%
    \else\y@ungabc(#1)%
    \fi}}\hskip\y@interspace}
\def\y@ungabc(#1,#2){%
  \y@nelineabcboxes{#1}%
  \y@lastargtest#2,
  \ify@islastarg\y@nelineabcboxes{#2}%
  \else\y@ungabc(#2)%
  \fi}
\catcode`\@12\relax
\def\punkt{\hbox{\vrule height10.5pt width 13.1pt depth 2.6pt}}
\def\Punkt{\hbox to 13pt{\vrule height10.5pt width 12.5pt depth 2.6pt\hfill}}
\def\leer{\hbox{}}

\renewcommand{\Bbb}{\mathbb}
\newcommand{\BZ}{\Bbb{Z}}

\newcommand{\tx}{\textsf{X}}
\newcommand{\toh}{\textsf{O}}
\newcommand{\be}{{\bf e}}
\newcommand{\dpp}{\operatorname{dp_+}}
\newcommand{\dpd}{\operatorname{dp}}
\newcommand{\SubjRand}{{\bf sr}}
\newcommand{\dset}{\operatorname{Dset}}

\newtheorem{Theo}{Theorem}

\begin{document}
\title{Effective Generation of Subjectively Random Binary Sequences}
\author{Yasmine B. Sanderson\footnote{Mathematisches Institut, Friedrich-Alexander-Universit\"{a}t, Erlangen-N\"{u}rnberg, Bismarckstra\ss e 1 1/2, 91054 Erlangen, Germany. Email: sanderson@mi.uni-erlangen.edu. This paper was written while at the Department of Mathematics, Rutgers University.}}
\date{}
\maketitle
\begin{abstract}
We present an algorithm for effectively generating binary sequences which would be
rated by people as highly likely to have been generated by a random process,
such as flipping a fair coin.
\end{abstract}
\section{Introduction}

This paper is a first step in  modelling mathematical objects showing 
``subjective randomness'', or what people believe to be
random. Although there is no rigorous characterization of what
subjective randomness might be, it has become clear through experimentation that
is quite different from stochastic randomness.  
A classic example which
illustrates this difference is the
following: when asked which of the following sequences is most likely be to produced by
flipping a fair coin 20 times,

\smallskip
\centerline{\textsf{OOOOOOXXXXOOOXXOOOOO}}
\medskip
\centerline{\textsf{OOXOXOOXXOOXOXXXOOXO}} 

\smallskip
\noindent
most people will answer ``the second
sequence'' even though each sequence has been produced by a random generator. 

 Until now, subjective randomness has mainly been the study of psychologists, 
cognitive scientists and artists\footnote{see Twyla Tharp's ``Medley'' or any 
of Jackson Pollock's drip paintings}. However, in today's age, where computer 
software is an integral part of everyday life, it is a natural problem to ask 
how one can present what people would accept as ``random''. Examples of this interest is the popularity of the various ``random'' playlist shuffling software on the market and the appearance of ``randomness'' in design (screensavers, tiling, etc.). In fact, this paper 
grew from an attempt to create exercise drill software for students. The goal 
was to generate images which would be simultaneously unpredictable and yet, in some way, 
balanced. We hope to be able to describe a good model for subjectively random 
two-dimensional objects sometime in the future. For now we present one for which 
there already exists a substantial body of research: subjectively random 
binary sequences.

Most of the research on subjective randomness seeks to understand 
what exactly are the differences between subjective randomness and stochastic 
randomness and to understand why this is
so. Experiments usually are of two types: ``production'' where
subjects are asked to produce examples of what they consider to be
random and ``judgment'' where subjects are asked to identify or
rate objects based on how likely they would have been produced by some
random process. From these experiments, a few traits of subjective
randomness have been pinpointed.

One is {\bf local representativeness} the fact
that people feel that a small sample should reflect properties of the
population as a whole [KT], [R1], [TK]. In terms of binary sequences, this translates
to concrete restrictions on the probability
 of alternation and the
relative number of \tx's and \toh's in any given subsequence. 

The second trait is high {\bf subjective complexity} or effort needed to
encode the data [C], [F2], [L3]. In the same way that algorithmic (Kolmogorov)
complexity is measured as the inability of a computer to encode the
sequence in less bits than the length of the sequence [LV], so can
perceived randomness be measured as the inability of the human mind to
memorize the sequence in significantly less steps than there are
sequence elements. 

The third trait concerns {\bf symmetry recognition}. People are less likely
to rate a sequence as random if they recognize certain symmetries in them [F3], [LO], [S], [R3], [T]. Which symmetries are recognized depends on whether the sequence elements are
presented simultaneously or one bit at at time.

In studying those sequences which are subject to the above constraints, it becomes 
apparent that subjectively random 
binary sequences are anything but random in the sense that they have properties which fall far from the norm for Bernoulli trials [LV]. This is also the reason that they could not be the output of any pseudo-random generator [K], [L1], [L2]; for one, these sequences only have very short runs of \tx's and \toh's. Although the locally even distribution of \tx's and \toh's in subjectively random sequences resembles that of low discrepancy, or quasi-random, sequences, they are not
quasi-random because the alternation rate of  \tx's and \toh's is too high from normal [CG].
 The structure of subjectively random sequences is quite rigid 
and these sequences constitute a very small subset of all binary sequences. In
fact, this set is so small (less than 10\% of all sequences of length
20 or more bits) that it makes more sense to use a creation method
than some brute force method using a random generator.
We use all of the above conditions in creating a simple algorithm
which generates (arbitrarily long) subjectively random binary sequences. We
concentrate on those sequences which would be presented one bit at a
time, such as in a ticker-tape or in the case when someone is actively
flipping a coin. 

Our algorithm depends on a function which measures subjective
randomness. This function is itself a variation on two ways of measuring
subjective randomness, one developed by Falk and Konold [FK], the
other developed by Griffiths and Tenenbaum [GT2] but is more efficient 
in measuring subjective randomness on
8-bit sequences. The sequence is created one bit at a time; at each
step one chooses the element which will produce the desired randomness
rating on the last 8 bits of the sequence. Producing this sequence is equivalent to moving along paths in a certain associated digraph. By traveling within a strongly connected digraph associated to sufficiently random 8-bit sequences, one can produce
 arbitrarily long binary sequences that consistently look ``random enough''. 

In the first few sections of the paper we give 
brief accounts of the research results which will be considered in
creating our subjectively random sequences. These sections restrict 
themselves to describing specific results and are not surveys of the 
substantial body of research in subjective randomness done by
psychologists and cognitive scientists. For that, we recommend
articles by Bar-Hillel and Wagenaar [BW], [W], Falk and Konold [FK] and by Nickerson [N]. 

We thank T. Griffiths for sharing data and answering questions and the referee for many useful comments and suggestions.

\section{Local Representativeness}

In their studies of how people perceived randomness,
Kahneman and Tversky argued that, generally, people believe that small
samples should reflect the properties of the population as a
whole. In probability theory, the Law of Large Numbers states that large samples tend to
have the same properties as the population as a whole. Kahneman
and Tversky coined the term ``The Law of Small Numbers'' to describe
people's belief that this holds even for small samples [KT]. When
small samples don't behave expectedly, people believe
that there must be another reason behind the data. For example,
gamblers attributed ``luck'' (good or bad) for streaks in outcomes [WK].

In the case of binary sequences the Law of Small Numbers presents
itself in the following way: people believe that in even short random
sequences there must be roughly the same number of \tx's as
there are \toh's with some irregularity in the
order of their appearance. In other words, neither {\it pure runs},
such as \textsf{XXXXXX} or \textsf{OOOOOO}, nor {\it alternating runs},
such as \textsf{XOXOXOXO} or \textsf{OXOXOXO}, should be too long. 
When considering long binary sequences, these same features should be
present in short subsequences, that is, in people's windows of
observation. These windows are believed to be of variable length of
approximately 7 bits [KG].

These conditions account for a high alternation rate in subjectively
random binary sequences. Truly random sequences tend to have an
alternation rate of 0.5, but studies consistently show that
subjectively random sequences have a probability of alternation of 0.6
- 0.7
[BW], [F1], [FK]. One way of understanding why this could be so is that, since 
there should consistently be
about as many \tx's and there are \toh's, if there is a run of \tx's, then
the run shouldn't be too long. This forces the probability of
alternation to be higher than average. On the flip side, the
restriction on the length of alternating runs forces the probability of
alternation to not become too high.

\section{Subjective Complexity}

In [FK], Falk and Konold find that people tend to view a sequence as
being more random if it was harder to mentally encode it. For example,
the sequence \textsf{XOXOXOXOXO} could be described as ``five times \textsf{XO}''. 
However, the sequence \textsf{XOOXXOXXOO} cannot be described so
concisely: 
``first an \tx, then two \toh's, then two \tx's, an \toh, then two \tx's, then two \toh's''.  
In order to objectively measure this effort, Falk and Konold define a 
``difficulty predictor'' $\dpd$. Let $x$ be a sequence. Then 
$$\dpd(x) = \mbox{the number of pure runs} + 2*\mbox{the number of
  alternating runs in $x$}$$
The idea behind this measure is that
one uses shortcuts such as ``four \tx's'' or ``\textsf{XO} five times'' to
describe the sequences. The difficulty predictor measures how many
different commands one would have to give, taking into account the
greater effort one needs to check the length of alternating runs over
pure runs (hence the factor 2). If there are multiple ways of
describing the sequence $x$ as concatenation of runs, then $\dpd(x)$ is
defined to be the minimum of the resulting numbers.

\smallskip
\noindent
{\bf Example:} Consider $x = $ \textsf{XXXOXOXOOO}. Then we can view $x$ as
$$\text{\underline{\textsf{XXX}}\ \underline{\toh}\ \underline{\tx}\ \underline{\toh}\ \underline{\tx}\ \underline{\textsf{OOO}} \ \  \mbox{or} \ \ 
\underline{\textsf{XXX}}\ \underline{\underline{\textsf{OXOX}}}\ \underline{\textsf{OOO}} \ \  \mbox{or} \ \ 
\underline{\textsf{XX}}\ \underline{\underline{\textsf{XOXOXO}}}\ \underline{\textsf{OO}}}$$
The first way rates $x$'s difficulty as 6, the second two rate it as 4. So $\dpd(x) = 4$.

In trials where subjects were asked to memorize and copy various
sequences of length 21, Falk and Konold found that their difficulty
predictor was a good correlator for the randomness rating that was
given by the subject and concluded that it was a good measure of the
sequence's perceived randomness [FK].

Consistent with previous studies, Falk and Konold also found that those
sequences rated as most random tended to have a higher than average probability of
alternation. 
It should be noted that this is, mathematically, not a
coincidence. Requiring a maximal value for $\dpd$
is a stronger condition than simply requiring that the probability of alternation 
be 0.6-0.7 and that the imbalance (defined as $|\mbox{sequence length}/2 -\#\tx's |$)
be minimal.

In a series of papers [GT1],[GT2],[GT3], Griffiths and Tenenbaum extend the
results of 
Falk and Konold with their own measure of subjective complexity. They propose that, when
presented with a 
binary sequence $x$, people assess the probability that $x$ is being
produced 
by a random process as opposed to being produced by some other (regular)
process. Their 
measure for subjective complexity is then $P(x| \text{regular})$,
(where $x$ is 
considered more random, the smaller $P(x| \text{regular})$ 
is).

 In
experiments where they ask subjects to order 8-bit sequences with
respect to how random they believe they are, Griffiths and Tenenbaum
first show that a good model for calculating $P(x| \text{regular})$ is a 
finite state automaton in the 
form of a certain hidden Markov model (HMM)[GT2].
Conceptually, the model works as follows: As someone reads a sequence $x$, 
they will consider each element, \toh \ or \tx, as possibly being part of one of 
several motifs. These 
motifs are of varying length and the probability that \toh \ or \tx \ will be considered 
as being part of a specific motif will be a function of the motif length and 
whether one is changing from one motif to another. 

\smallskip
 {\bf Example:} In the smallest HMM model they consider [GT2], the 6 state model, there
 are the following states which produce an \toh \ as observed occurrence: \toh \ 
 (coming from the motif \toh), \toh \  (coming from the motif \textsf{OX}) and \toh \ (coming from the motif
 \textsf{XO}). Similarly, there are 3 states which produce \tx. The probability
 of remaining in a motif is denoted by $\delta$, the probability of
 changing to (or starting at) a motif of length $k$ is equal to 
$C \cdot \alpha^k$ (for some $\alpha$) where 
$C := (1-\delta)/(2\alpha + 2\alpha^2)$ is a normalization constant. 

Let {\bf 1} denote the state which produces \tx \ (from the length 1 motif
\tx), {\bf 2} denote the state which produces \toh \ (from the motif \toh), {\bf 3},
resp. {\bf 4}, denote the state which produces \tx, resp. \toh, from the
motif \textsf{XO}, and {\bf 5}, resp. {\bf 6}, denote the state which produces \tx,
resp. \toh, from the motif \textsf{OX}. The transition matrix giving $P({\bf i}|{\bf j})$ of
going from state {\bf i} to state {\bf j} is:
$$P({\bf i}|{\bf j}) = \left( \begin{array}{cccccc}
\delta & C\alpha & C\alpha^2 & 0 & 0 & C\alpha^2 \\
C\alpha & \delta & C\alpha^2 & 0 & 0 & C\alpha^2 \\
C\alpha & C\alpha & 0  & \delta & 0 & C\alpha^2 \\
C\alpha & C\alpha  & \delta & 0   & 0 & C\alpha^2 \\
C\alpha & C\alpha   & C\alpha^2  & 0   & 0& \delta \\
C\alpha & C\alpha   & C\alpha^2  & 0   & \delta & 0 
\end{array} \right)$$

In the same spirit as for the difficulty predictor, there are many 
different sequences of states which can produce the same observed sequence. 
For example, the sequence $x = $ \textsf{XOXX} could be produced by any of many
state sequences: the first \tx \  could come from state {\bf 1} or {\bf 3}, the
\toh \ could come from states {\bf 2}, {\bf 6} or from state {\bf 4} if the preceding
\tx \ was from state {\bf 3}, the second \tx \ could come from state {\bf 1} or
{\bf 3} or from state {\bf 5} if the preceding \toh \ was from state {\bf 6}, and the
third \tx \ could come from state {\bf 1} or {\bf 3} (if the previous \tx \ was not
also from state {\bf 3}). For each state sequence
$z$, the probability $P(x,z)$ that the observed sequence $x$ was produced from $z$
 will be a function of $\delta$ and $\alpha$. For example
 $P(\text{XOXX},{\bf 1211}) = (C\alpha)^4 = C^4\alpha^4$ and
 $P(\text{XOXX},{\bf 3431}) = C\alpha^2 \delta^2 C\alpha =
 C^2\alpha^3\delta^2$. The probability
$P(x|\text{regular})$ is defined as the $\max_z P(x,z)$ over all possible state
sequences which produce $x$. 
\smallskip

The 22 state model is the natural extension of the 6 state model to 
include up to length $4$ motifs which are not duplications of smaller 
length motifs (such as \textsf{XOXO} or \textsf{XXXX}). 
\smallskip

Both the 6 state and the 22 state models have the advantage over the
difficulty predictor of being a function of the length of a
sequence.
For certain values of $\delta$ and $\alpha$, Griffiths and Tenenbaum
show that there is a equivalence of the 6 state model and Falk and Konold's difficulty
predictor. In addition, they find values of $\delta$ and
$\alpha$ such that 
both HMMs modelled the trials' subjective randomness results
better than the difficulty predictor [GT2]. As with the difficulty predictor, 
being rated maximally random (by the 22 state model) is a stronger condition 
than requiring that the sequence has probability of alternation 0.6-0.7 and 
that the imbalance between \toh's and \tx's is low.

\section{Symmetry recognition}

Griffiths and Tenenbaum's experiments also test the role of symmetry
recognition in subjective 
randomness and consider the possibility that
subjects recognize four types of symmetry: mirror symmetry, where the
second half is produced by reflection of the first half (ex. \textsf{XXOOOOXX}),
complement symmetry, where  the
second half is produced by reflection of the first half and exchanging
\tx \ and \toh \ (ex. \textsf{XOOXOXXO}), and duplication, where the sequence is
produced by repeating the first half once (ex. \textsf{XXOXXXOX}). 
Their ``context-sensitive'' model
for $P(x|\text{regular})$ considers that sequences be generated by any of four
methods. The first method is repetition, where sequences are produced
by the HMM. The next three methods are the symmetry methods listed
above, where the first half of the sequence is produced by the HMM and
the second half by the symmetry rule.
Then, $P(x|\text{regular})$ will
depend on the models $M$ listed above: $$P(x|\text{regular}) =
\max_{z,M}P(x,z|M)P(M)$$ where $P(x,z|M)$ is obtained as above and
$P(M)$ is to be determined by analysis of hard data.

In [GT3], they conclude that when all elements of a sequence are
presented simultaneously, their context-sensitive model models
subjective randomness best with the following parameters: $\delta =
0.66$, $\alpha = 0.12$, $P(\text{repetition}) = 0.748$, $P(\text{mirror symmetry})=
0.208$, $P(\text{complement symmetry}) = 0.0005$ and $P(\text{duplication})=0.039$.
However, when elements of a sequence are presented sequentially,
people do not recognize mirror or complement symmetry, so the best
model has the following parameters $\delta =
0.70$, $\alpha = 0.11$, $P(\text{repetition}) = 0.962$, and
$P(\text{duplication})=0.038$.

\section{Measuring subjective randomness effectively}

Griffiths and Tenenbaum find values for $\alpha$ and $\delta$ in order to fit their model to the linear ordering on the sequences which is obtained by the 
 experimental data. However,
for our problem we do not need the full structure of their model; since we are interested in the sequences rated ``more random'', we only need the linear ordering of the sequences 
produced by this
model. An analysis of the HMM model shows that we obtain the same ordering as given by the 22-motif HMM by doing the following. We consider $\alpha$ and $\delta$ as formal
parameters satisfying $0 < \alpha < \delta^4 < 1$. The normalizer $C$ can be set to $C=1$.  This abstract version of the HMM will assign
to each sequence an expression of the form $\alpha^a\delta^b$. Let
$x$ and $y$ be two sequences such that $P(x|\text{regular}) =
\alpha^{i_1}\delta^{j_1}$ and $P(y|\text{regular}) = \alpha^{i_2}\delta^{j_2}$ (where
the probability $P( \cdot |\text{regular})$ is determined by the
HMM). Then, $x$ is subjectively more random than $y$ if $i_2-i_1 \geq 0$
and $j_1 - j_2 \leq 4(i_2 - i_1)$\footnote{This does not give a partial ordering on {\it all} monomials in $\alpha$ and $\delta$ but simply on those produced by this model.}.

 With regard to the repetition and duplication models, we need to set 
$ \alpha\delta^4 <
P(\text{duplication})/P(\text{repetition}) < \alpha\delta^3$. To
obtain the same ordering as in [GT3], it suffices to set
$P(\text{repetition})=1$ and $P(\text{duplication})=\alpha\delta^{3.5}$.
(In our next paper, we
will consider the simultaneous case which includes all four models).

Table 1 gives the partial ordering of all 128 sequences
starting with \toh \  from most to least subjectively random.The first
column gives an example sequence in full and the second column gives
the list of equally rated sequences in the form of a base 10
number. 
The third column gives the ``finite state'' $P(x|\text{regular})$ as determined by
the 22 motif HMM. The fourth column gives the ``context-sensitive''
$P(x|\text{regular})$ which also takes into consideration duplication symmetry.

\begin{table}[t]
\centering
\begin{tabular}{|p{4 cm}|p{6 cm}|p{1.5 cm}|p{1.5 cm}|}
\hline sample sequence & all sequences with same randomness rating & finite state & context-sensitive\\
\hline [0, 1, 0, 0, 1, 1, 0, 1] & 77 & $\alpha^5\delta^6$ & $\alpha^5\delta^6 $ \\ 
\hline [0, 1, 1, 0, 1, 0, 0, 1] & 105 & $\alpha^5\delta^5$ & $\alpha^5\delta^5 $ \\ 
\hline   [0, 1, 1, 0, 1, 0, 1, 1] & 41, 69, 74, 82, 89, 93, 101, 107, & $\alpha^5\delta^4$ & $\alpha^5\delta^4 $ \\ 
\hline [0, 1, 1, 1, 0, 1, 1, 0]  & 38, 44, 46, 50, 52, 66, 70, 76, 78, 98, 100, 110, 114, 116, 118 & $\alpha^5\delta^3$ & $\alpha^5\delta^3 $ \\ 
\hline [0, 1, 1, 0, 1, 1, 0, 0]  & 18, 22, 37, 45, 54, 72, 75, 90, 91, 104, 108 & $\alpha^4\delta^6$ & $\alpha^4\delta^6 $ \\ 
\hline  [0, 1, 1, 1, 1, 0, 1, 0] & 20, 26, 40, 43, 53, 58, 81, 83, 86,
88, 92, 94, 106, 117, 122 & $\alpha^4\delta^5$ & $\alpha^4\delta^5 $
\\ 
\hline [0, 1, 1, 0, 0, 1, 1, 0]  & 34, 68, 102 & $\alpha^4\delta^7$ & $\alpha\delta^{4.5} $ \\ 
\hline [0, 1, 1, 1, 1, 1, 0, 1]  & 9, 11, 13, 19, 23, 25, 27, 29, 33, 35, 39, 47, 49, 55, 57, 
  59, 61, 65, 67, 71, 79, 97, 99, 103, 111, 113, 115, 121, 123, 125 & $\alpha^4\delta^4$ & $\alpha^4\delta^4 $ \\ 
\hline [0, 1, 1, 0, 1, 1, 0, 1]  & 36, 73, 109 & $\alpha^3\delta^7$ & $\alpha^3\delta^7 $ \\
\hline  [0, 1, 0, 1, 1, 1, 1, 1] & 5, 10, 21, 42, 80, 84, 87, 95 & $\alpha^3\delta^6$ & $\alpha^3\delta^6 $ \\ 
\hline  [0, 1, 1, 1, 0, 1, 1, 1] & 17, 51, 119 & $\alpha^4\delta^4$ & $\alpha^3\delta^{5.5}$ \\ 
\hline [0, 1, 1, 1, 0, 0, 0, 0]  & 2, 4, 6, 8, 12, 14, 16, 24, 28, 30, 32, 48, 56, 60, 62, 64, 96, 112, 120, 124, 126 & $\alpha^3\delta^5$ & $\alpha^3\delta^5 $ \\ 
\hline  [0, 1, 0, 1, 0, 1, 0, 1] & 85  & $\alpha^2\delta^7$ & $\alpha^2\delta^7 $ \\ 
\hline [0, 1, 1, 1, 1, 1, 1, 1]  & 1, 3, 7, 15, 31, 63, 127 & $\alpha^2\delta^6$ & $\alpha^2\delta^6 $ \\ 
\hline  [0, 0, 0, 0, 0, 0, 0, 0] & 0 & $\alpha^1\delta^7$ & $\alpha^1\delta^7 $ \\ \hline
\end{tabular}
\caption{The linear ordering of 8-bit binary sequences given by both
``finite state'' and ``context-sensitive'' models. Sequences $x_1\cdots x_n$ are written in base 10: $2^{n-1}x_1 + \cdots + 2x_{n-1} + x_n$}
\end{table}

This abstract version of the 22-motif HMM allows us to define a more efficient variant $\dpp$, which will give the same result as the 22-motif HMM; if the HMM assigns a sequence $x$ the value $\alpha^i\delta^j$
then this predictor will assign it the value $[i,j]$, where $i$
represents the sum of the lengths of the contributing motifs and the $j$ represents
the length of the sequence minus the number of contributing motifs. However, it will ``read'' the sequence 
in the same spirit as Falk and Konold's difficulty predictor. In addition to 
length 1 and 2 motifs, 
$\dpp$ considers length 3 and 4 motifs and records 
over how many bits the motif appears. For example, viewing the
sequence \textsf{XOOXOOXX} as two instances of \textsf{XOO} and two
instances of \tx\  gives a
value of $[3 + 1,8-2] = [4,6]$  Here, we
allow incomplete repeats to be included into the run.  For example, the
sequence \textsf{XOXOXOX} would be given a value of $[2,6]$
because it is a run of 3 1/2 instances of the length 2 motif \textsf{XO}. 
The rating given to a sequence $x$ would be the smallest
$[i,j]$ obtained by considering $x$ as various concatenations of motifs. 
The part of $\dpp$
consisting of the initial conditions and Step 1 is equivalent to
the abstract 6 state HMM (with conditions $C=1$ and $0<a<\delta^2
<1$). The reason that $\dpp$ is more efficient (for this problem) than 
the HMM is that it identifies those subsequences (and motifs) 
which will contribute to the value of $\dpp$. Hence, over 8-bit sequences, 
it is recursive with small depth. For any object $c$ described using
bracket notation $[ \cdots ]$, we 
will write $c[i]$ to denote its $i$th component or entry.
The algorithm for $\dpp$ is as follows:

\medskip \noindent
{\tt{\bf Algorithm to compute $\dpp$:} 
 
   INPUT: $x = x_1 x_2 \cdots x_n$.
 
\smallskip
\noindent
\# Special cases:
  
   IF $x$ is empty THEN OUTPUT $[0,0]$ 

   ELSE IF $x$ is a pure run THEN OUTPUT $[1,n-1]$

   ELSE IF $x$ is an alternating run and $n \geq 3$ THEN OUTPUT $[2,n-1]$ 

\medskip \noindent
\# consider only length 1 and 2 motifs

   ELSE SET $j:=1$

\# find the largest $j\geq 1$ such that $x_1 \cdots x_{j}$ consists of
$j$ repeats of $x_1$.

\hskip 1cm WHILE $j \leq n$ AND $x_j = x_1$ DO

\hskip 1.5cm INCREMENT $j$ BY 1

\hskip 1cm END DO

 \hskip 1cm  IF $j > 1$ THEN

  \hskip 1.5cm $s1 := x_1 x_2 \cdots x_{j-1}$ and $s2 :=  x_{j} \cdots x_n$.

   \hskip 1cm ELSE SET $k:=1$

\# find the largest $k\geq 1$ such that $x_1 x_2 \cdots x_{2k}$ 
is $k$ repeats of $x_1x_2$.

\hskip 1cm WHILE $k \leq n$ AND $x_{2k-1} = x_1$ AND $x_{2k}=x_2$ DO

\hskip 1.5cm INCREMENT $k$ BY 1

\hskip 1cm END DO

   \hskip 1.5cm IF $k=1$ THEN 
   
\hskip 2cm $s1 := x_1 $ and $s2 :=  x_{2} \cdots x_n$.

   \hskip 1.5cm ELSE $s1 := x_1 x_2 \cdots x_{2k}$ and $s2 :=  x_{2k+1} \cdots x_n$. 

   \hskip 1.5cm END IF

\hskip 1cm END IF

END IF

   SET $\dpp^{\prime}(x) = \dpp(s1) + \dpp(s2)$.

\medskip
\noindent
\# consider length 3 motifs 

 SET $\dset = \{\dpp^{\prime}(x)\}$ \# collects possibly smaller difficulty values

 IF $\dpp^{\prime}(x)[1] \geq 4$ THEN

\hskip 1cm FOR $i$ FROM $1$ TO $ n-5$ DO

\hskip 1.5cm  \# determine the largest $j\geq i+4$ such that $x_i...x_j$ is a repeat
   of $x_i x_{i+1} x_{i+2}$. 

\hskip 1.5cm SET $j:=i+3$

\hskip 1.5cm WHILE $j \leq n$ AND $x_j = x_{i + ((j-i) \mod 3)}$ DO

\hskip 2cm INCREMENT $j$ BY 1

\hskip 1.5cm END DO

 \hskip 2cm IF $j-i \geq 4$ THEN

\hskip 2.5cm LET $\dset := \dset $ UNION $\{\dpp(x_1 \cdots x_{i-1}) + [3,j-i-2] +
   \dpp(x_{j+1} \cdots x_n)\}$

\hskip 2cm END IF

\hskip 1cm END LOOP

END IF

SET $\dpp^{\prime\prime}(x)= \min_{v\in \dset} v $

\medskip
\noindent
\#consider length 4 motifs

 SET $\dset = \{\dpp^{\prime\prime}(x)\}$ \# collects possibly smaller difficulty values

 IF $\dpp^{\prime\prime}(x)[1] \geq 5$ THEN

\hskip 1cm FOR $i$ FROM $1$ TO $ n-6$ DO

  \hskip 1.5cm   \# determine the largest $j\geq i+5$ such that $x_i...x_j$ is a repeat
   of $x_i x_{i+1} x_{i+2} x_{i+3}$. 
  
  \hskip 1.5cm SET $j:=i+4$
 
  \hskip 1.5cm WHILE $j \leq n$ AND $x_j = x_{i + ((j-i) \mod 4)}$ DO

\hskip 2cm INCREMENT $j$ BY 1

\hskip 1.5cm END DO

 \hskip 2cm IF  $j-i \geq 7$ THEN

\hskip 2.5cm LET $\dset := \dset$ UNION $\{\dpp(x_1 \cdots x_{i-1}) + [4,j-i-2] +
   \dpp(x_{j+1} \cdots x_n)\}$

\hskip 2cm END IF

\hskip 1cm END LOOP

END IF 

OUTPUT: $\min_{v\in \dset} v $

END IF

}

\medskip

Using MAPLE, we found that, $\dpp$ took significantly less time and memory than the 
(polynomial time) Viterbi algorithm (see [R2] for a good tutorial on
the Viterbi algorithm) with the 22 motif HMM. On average, the Viterbi algorithm takes 47 
times the time and 400 times the space as $\dpp$ to calculate the difficulty of an 
8-bit sequence. We also note that this algorithm duplicates the results of the 22 
motif model only for sequences of 8 bits or less. It becomes less accurate as the 
length of the sequence increases. 

\medskip
Taking into account the possibility of duplication ($x= x_1 x_2 \cdots x_{n/2}x_1 x_2 \cdots x_{n/2}$), 
we obtain the subjective randomness rating  $\SubjRand(x)$ for any 8-bit sequence $x$:
$$\SubjRand(x) = \begin{cases} \min(\dpp(x) , \dpp(x_1 x_2
\cdots x_{n/2})+[1,3.5]) &  \text{if $x$ is a perfect duplication,}\\ 

\dpp(x) & \text{otherwise}\end{cases}$$

This gives the same partial ordering on 8-bit sequences as the "context-sensitive" model 
described in Section 3.

\section{Sequence Creation}

We can now produce arbitrarily long subjectively random binary
sequences. In this paper, we concentrate on those sequences which
would be presented sequentially (one bit at a time) instead of
simultaneously. According to [GT3] (see above), in addition to the
rating given to the sequence by the HMM, the only symmetry that
needs to be considered then is duplication.

The idea behind producing this sequence is the obvious one: pick the
 next element in a sequence (\toh \ or \tx) to be the one which gives a
 sufficiently high rating of 
 subjective randomness on the last 8 bits of the resulting sequence.
 The rationale behind this is that since people use small (6-8
 bits) windows of
 observation when analyzing a sequence, it suffixes that people
 feel at any one time
 that their window looks ``random'' enough.

Ideally, one would like to create sequences which would be of maximal
$\SubjRand$ over every small subsequence. However, such
sequences simply do not exist. Even restricting oneself to sequences
for which every 8-bit subsequence $x$ is in the top 20 \%
($\SubjRand(x)[1]=5$) becomes too
deterministic; there are only 2 such
(arbitrarily long) sequences and each has a period of 6 bits. 

 For any 8-bit sequence $x=x_1\cdots x_8$, let $x\toh:=x_2\cdots x_8$\toh\
and $x\tx:=x_2\cdots x_8$\tx. (If $x$ is the last 8 bits of the sequence
that we are creating, then $x\toh$ or $x\tx$ will be the last 8 bits
of the sequence to which we've added one more element).
To any subset $S$ of 8-bit sequences, we can associate a directed graph, or digraph, $G:=G(S)$ in the following way: the vertices of $G$ are indexed by the elements of $S$. The directed edges are defined as follows: 
$x \rightarrow y$ if $y=x\toh \in S$ or $y=x\tx \in S$.

Recall that a digraph $G$ is {\it connected}, resp. {\it strongly connected}, if, for every two vertices $x\ne y$ in $G$, there is a nondirected, resp. directed, path from $x$ to $y$. A subgraph $C$ of $G$ is a connected, resp. strongly connected, {\it component} if it is maximal for this property. A component is {\it trivial} if it consists of a single vertex. For our problem, we are particularly interested when $G$ is strongly connected. In this case, we can form arbitrarily long sequences $x$ by choosing an equally long directed path in $G$. The strong connectivity implies that there are no sinks and thus guarantees the existence of such a path. 

Set $S([m,n]) := \{x | x \mbox{ 8-bit }, \SubjRand(x)\geq [m,n] \}$ for $[m,n]$ any of the possible $\SubjRand$ values $[1,7]$, $[2,6], \ldots$,$[5,6] $. The following theorem is proved by brute force (Maple).
\begin{Theo} 1) For $[m,n] < [5,3]$, $G(S([m,n])) $ has exactly one non-trivial strongly connected component $C([m,n])$ and $C([5,3])$ has exactly two non-trivial strongly connected components $C_1([5,3])$, $C_2([5,3])$. These components form a sequence of nested digraphs:
$$C([1,7]) \supset  C([2,6]) \supset \cdots \supset C([4,6]) \supset C_i([5,3]), \ \ \ i=1,2.$$
$G(S([m,n])) $ has no non-trivial strongly connected components for $[m,n] > [5,3]$. 

2) For $[m,n] \leq [4,4.5]$, the $C([m,n])$ are also connected components of $G(S([m,n])) $.
\end{Theo}
Table 2 gives detailed information about the $C([m,n])$. A highly efficient way to produce sequences would be to use an incidence matrix (or table) $M$ for the graph $C([m,n])$, indexed by its vertices, with entries $M_{x,y}$ equalling the probability of traveling from $x$
to $y$.  A sequence would then be created by randomly choosing 
a vertex of $C([m,n])$ 
(which gives the first 8 bits of the sequence) and then moving to each 
consecutive vertex/bit as dictated by $M$. This method is fast (a million bits in 347 seconds using Maple) and it produces every possible sequence. However it is rigid; if one wants to tinker with the set from which these subsequences come (by making it larger), it would be preferable to use
a method that it based on built-in bounds. The success of such a method would depend on being always able to remain within a subgraph which is connected and strongly connected. The subgraph being connected allows one to use the algorithm's simple bounds condition and still remain in the graph. The strong connectedness prevents one from travelling to a sink. The algorithm is then straightforward:

\medskip
\noindent {\bf Sequence Creation Algorithm:} Input an initial $x=x_1 \cdots x_8$ from $C([m,n])$ and $N$, the desired sequence length. For $i$ from $9$ to $N$ do: 

$\bullet$ if $\SubjRand(x\tx) \geq [m,n]$ AND $\SubjRand(x\toh) < [m,n]$ then set $x_i = \tx$ 

$\bullet$ if $\SubjRand(x\toh) \geq [m,n]$ AND $\SubjRand(x\tx) < [m,n]$ then set $x_i = \toh$ 

$\bullet$ otherwise choose $x_{i}=$ \toh \  or \tx \ at random.

\medskip
From the above Theorem, we see that we can use this algorithm for paths within $C([m,n])$, $[m,n] \leq [4,4.5]$. (Otherwise, we need more complicated conditions to ensure that we remain within $C([m,n])$.)
We argue that, for very long sequences, the best choice is the 164-element set $C([4,4])$.
This might at first seem to be too large of a set. However, any
20-bit sequence produced from $C([4,4])$ is in the top 9\% (of subjective randomness) of all
20-bit (or more) sequences. $C([4,4])$ also has a large cycle basis which allows for a great variety of sequences (see [GLS] for results concerning cycle bases for digraphs). Since the proportion of \# cycles/ \# vertices is significantly larger for $C([m,n])$ (see Table 2) than it is for any $C([m,n])$ for $[m,n]>[4,4]$ it makes sense to choose $C([4,4])$ to make these sequences.
 Almost any 7-bit sequence can serve as the first seven bits of an element in $C([4,4])$. The exceptions are 1) any sequence with a streak of length 5 or more, 2) those 7-bit sequences with $\SubjRand [1] \leq 2$, and 3) the sequences \textsf{OOOXOOO} and \textsf{XXXOXXX}. With such an initial sequence and bounds $[m,n] = [4,4]$ in the above algorithm, we produce sequences always within $C([4,4])$.

\medskip
On average, such subjectively random sequences coming from $C([4,4])$ have equal numbers of \tx's and \toh's and an alternation rate of .58. Examples of such subjectively random sequences coming from are:
$$\young(\punkt\leer\punkt\leer\leer\leer\punkt\leer\leer\Punkt\punkt\leer\punkt\leer\leer\Punkt\Punkt\punkt\leer\leer\punkt\leer\leer\leer\leer\Punkt\punkt\leer\punkt\leer)
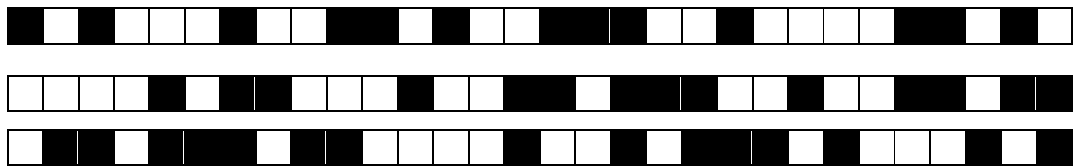
$$
$$\young(\leer\leer\leer\leer\punkt\leer\Punkt\punkt\leer\leer\leer\punkt\leer\leer\Punkt\punkt\leer\Punkt\Punkt\punkt\leer\leer\punkt\leer\leer\Punkt\punkt\leer\Punkt\punkt)$$
$$\young(\leer\Punkt\punkt\leer\Punkt\Punkt\punkt\leer\Punkt\punkt\leer\leer\leer\leer\punkt\leer\leer\punkt\leer\Punkt\Punkt\punkt\leer\punkt\leer\leer\leer\punkt\leer\punkt)$$
which is consistently more ``balanced'' than what is produced by a random generator:
$$\young(\punkt\leer\punkt\leer\Punkt\Punkt\punkt\leer\Punkt\Punkt\punkt\leer\punkt\leer\leer\Punkt\punkt\leer\leer\leer\leer\leer\leer\leer\leer\Punkt\punkt\leer\punkt\leer)
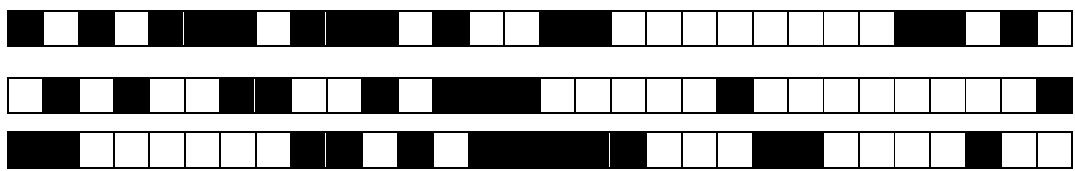
$$
$$\young(\leer\punkt\leer\punkt\leer\leer\Punkt\punkt\leer\leer\punkt\leer\Punkt\Punkt\punkt\leer\leer\leer\leer\leer\punkt\leer\leer\leer\leer\leer\leer\leer\leer\punkt)$$
$$\young(\Punkt\punkt\leer\leer\leer\leer\leer\leer\Punkt\punkt\leer\punkt\leer\Punkt\Punkt\Punkt\Punkt\punkt\leer\leer\leer\Punkt\punkt\leer\leer\leer\leer\punkt\leer\leer)$$

\section{Conclusion}
There are many questions that one might ask concerning the sequences presented here as subjectively random. 
For example, wouldn't people notice periodicity with motifs of length 5 or more? 
Would a HMM with more motifs give a more accurate model for subjective randomness? Would it be better to use 
larger ``windows'' when evaluating a sequence's subjective randomness? How do the conditions for subjective randomness change as sequences become longer?
Would these sequences be, in the long run, {\it too} regular (and thereby too predictable)? 

These and similar questions bring us to the main problem for this program, that is, the lack of hard results 
which are geared to this type of modeling. 
Simply put, most research done on subjective randomness seeks to answer questions which have little to do 
with producing subjectively random objects. 
The algorithm presented here is based on what information there is. However, more research would have to be 
done to determine how to fine-tune it, if necessary. 

\medskip\noindent
\begin{table}[thb]
{
\begin{tabular}{|p{1.5 cm}|p{7 cm}|p{1.75 cm}|p{1.25 cm}|p{2 cm}|}
\hline $[m,n]$ & $C([m,n])$ & num. vertices & num. arcs & cycle basis cardinality\\
\hline
 $[5,3]$ & $C_1([5,3])=\{ 44, 89, 178, 101, 203, 150\}$ & 6 & 6 & 1 \\  
& $C_2([5,3])=\{77, 154, 52, 105, 211, 166\}$  & 6 & 6 & 1 \\
\hline $[4,6]$ & $C_1([5,3])\ \cup \ C_2([5,3]) \ \cup $ & 46 & 58 & 13 \\
  & $\{ 22, 38, 41, 45, 46, 54, 69, 74, 75, $ & & & \\
& $82, 90, 93, 100, 104, 107, 108, 116,  139,$ & & & \\
  & $ 147, 148, 151, 155, 162, 165, 173, 180,$ & & & \\
 & $ 181, 186, 201, 209, 210, 214, 217, 233\}$ & & & \\
\hline $[4,5]$ & $S([4,5]) \setminus \{18, 37, 50, 66, 72, 76, 91, 94,$ &80 &120 & 41\\
   &  $ 110, 118, 122, 133, 137, 145, 161, 164, 179,$&&& \\&$  183, 189, 205, 218, 237\}$ &&& \\ 
\hline $[4,4.5]$ & $S([4,4.5]) \setminus \{ 66, 94, 122, 133, 161, 189\}$ & 102 & 158 & 57 \\
\hline $[4,4]$ & $S([4,4]) \setminus \{190,125,65,130 \}$ & 164 & 280 & 117 \\
\hline $[3,7]$ & $S([3,7]) \setminus \{190,125,65,130 \}$ &170 & 298 & 129 \\
\hline $[3,6]$ & $S([3,6])$ &190 & 342 & 153 \\
\hline $[3,5.5]$ & $S([3,5.5])$ & 196 & 360 &165  \\
\hline $[3,5]$ & $S([3,5])$ & 238 & 462 & 225 \\
\hline $[2,7]$ & $S([2,7])$ & 240 & 467 & 228 \\
\hline $[2,6]$ & $S([2,6])$ & 254& 505 &  252\\
\hline $[1,7]$ & $S([1,7])$ = all 8-bit sequences & 256 & 512 & 254 \\
\hline
\end{tabular}
}
\caption{Description of the strongly connected components $C([m,n])$ for each $[m,n]$ for which $C([m,n])$ is non-trivial. 
(Sequences $x_1\cdots x_n$ are written in base 10: $2^{n-1}x_1 + \cdots + 2x_{n-1} + x_n$)}
\end{table}

\eject\noindent
\centerline{{\bf \Large References}}

\medskip
[BW] M. Bar-Hillel and W. Wagenaar, The Perception of Randomness,
 {\it Adv. in Applied Math.} {\bf 12} (1991) 428-454.

[C] N. Chater,  Reconciling simplicity and likelihood principles in perceptual organization. {\it Psych. Rev.} (1996) 103:566-581.

[CG] F.R.K. Chung and R.L. Graham, Quasi-random subsets of $\BZ_n$., {\it J. Combin. Theory} {\bf Ser. A 61.} (1992), no. 1, 64–86.
 
[F1] R. Falk, The perception of randomness, {\it Proceedings, Fifth International Conference for the Psychology of Mathematical Education, Grenoble, France} (1981) 222-229.

[F2] J. Feldman,  Minimization of Boolean complexity in human concept learning. {\it Nature}, {\bf 407}, (2000), 630-633.

[F3] J. Feldman, How surprising is a simple pattern? Quantifying ``Eureka!'', {\it Cognition}, {\bf 93}, (2004), 199-224.

[FK] R. Falk and C. Konold, Making Sense of Randomness: Implicit
Encoding as a Basis for Judgment, {\it Psychological Review} {\bf
  104} {\bf No. 2} (1997) 301-318.

[GLS] P.M. Gleiss, J. Leydold, and P.F. Stadler, Circuit Bases of Strongly Connected Digraphs, {\it Disc. Math. Graph Theory}, {\bf 23} {\bf Part 2} (2003) 241-260.

[GT1] T. Griffiths and J. Tenenbaum, Randomness and Coincidences: Reconciling Intuition and Probability Theory, {\it 23rd Annual Conference of the Cognitive Science Society} (2001) 370-375.

[GT2]  T. Griffiths and J. Tenenbaum, Probability, algorithmic
complexity, and subjective randomness, {\it Proceedings of the Twenty-Fifth Annual Conference of the Cognitive Science Society} (2003).

[GT3] T. Griffiths and J. Tenenbaum, From Algorithmic to Subjective
Randomness, {\it Advances in Neural Information Processing Systems 16} (2004).

[K] D. E. Knuth. The Art of Computer Programming, volume 2: Seminumerical Algorithms. Addison-Wesley, Reading, MA, second edition, 1981. 

[KG] M. Kubovy and D.L.Gilden, Apparent randomness is not always the
complement of apparent order, in ``The Perception of Structure''
(G. Lockhead and J.R. Pomerantz, Eds.), Amer. Psychol. Assoc.,
Washington, DC, 1990.

[KT] D. Kahneman and A. Tversky, Subjective probability:
A judgment of representativeness. {\it Cognit. Psych.}, (1972) 3:430-454.

[L1]  J.C. Lagarias, Pseudorandom Numbers. {\it Statistical Science}, {\bf 8} (1993) 31-39. 

[L2] P. L'Ecuyer, Random number generation from {\it Handbook of Computational Statistics}, J. E. Gentle, W. Haerdle, and Y. Mori, eds., Springer-Verlag, 2004. 

[L3] E.L.L. Leeuwenberg, Quantitative specification of information in sequential patterns. {\it Psych. Rev.} (1969), 76:216-220.

[LO] L.L. Lopes and G.D. Oden, Distinguishing between random and nonrandom events. {\it J. Exp. Psych: Learning, Memory and Cognition} (1987) 13:392-400.

[V] M. Li and P. Vitanyi, {\it An introduction to Kolmogorov complexity and its applications}, New York: Springer (1993).

[N] R. Nickerson, The Production and Perception of Randomness, {\it
  Psych. Rev.} {\bf 109} {\bf No. 2} (2002) 330-357.

[R1] M. Rabin, Inference By Believers in the Law of Small Numbers, {\it Quart. J. Econ.}, {\bf  117}, {\bf No. 3}, (2002) 775-816.

[R2] L. R. Rabiner, A Tutorial on Hidden Markov Models and Selected
Applications in Speech Recognition, {\it Proc. of the IEEE} {\bf 77}
{\bf No. 2} (1989) 257-286.

[R3] F. Restle. Theory of serial pattern learning, {\it Psych. Rev}, (1970), {\bf 77:}369-382.

[S] H.A. Simon, Complexity and the representation of patterned sequences of symbols. {\it Psych. Rev.}  {\bf 79} (1972) 369-382.

[TK] A. Tversky and D. Kahneman, Belief in the Law of Small Numbers,
{\it Psych. Bull.}, {\bf 76} {\bf No. 2} (1971) 105-110.

[W]  W. A. Wagenaar,, Generation of random sequences by human subjects: A critical survey of literature {\it Psych. Bull} {\bf Vol. 77}, {\bf No. 1}, (1972) 65-72.

[WK] W. A. Wagenaar and G. B. Keren, Chance and luck are not the
same. {\it J. Behavioral Decision Making}, {\bf 1} (1988) 65-75.

\end{document}